\documentclass[9pt,twocolumn,twoside]{osajnl}
\usepackage{graphicx}
\usepackage{amsmath,amsthm,amssymb}
\usepackage{array}
\usepackage{hyperref}
\usepackage{xcolor}
\usepackage{mathtools}
\usepackage{bm,times,enumitem}
\newcommand{\openone}{\leavevmode\hbox{\small1\normalsize\kern-.33em1}}

\newcommand{\D}{\mathrm{d}}
\newcommand{\I}{\mathrm{i}}
\newcommand{\RE}{\mathrm{Re}}

\newcommand{\ket}[1]{|{#1}\rangle}
\newcommand{\bra}[1]{\langle{#1}|}
\newcommand{\TP}[1]{{#1}^\textsc{t}}
\newcommand{\rvec}[1]{\bm{#1}}
\newcommand{\dyadic}[1]{\bm{#1}}
\newcommand{\tr}[1]{\mathrm{tr}\{#1\}}

\newcommand{\inner}[2]{\langle{#1}|{#2}\rangle}
\newcommand{\opinner}[3]{\langle{#1}|{#2}|{#3}\rangle}
\newcommand{\E}[1]{\mathrm{e}^{\mbox{\footnotesize$#1$}}}
\newcommand{\HERM}[2]{\mathrm{H}_{\,#1}\!\left(#2\right)}
\newcommand{\LAG}[2]{\mathrm{L}_{\,#1}\!\left(#2\right)}
\newcommand{\ALAG}[3]{\mathrm{L}^{(#1)}_{\,#2}\!\left(#3\right)}
\DeclareMathOperator{\sinc}{sinc}

\journal{optica} 

\setboolean{shortarticle}{false}

\title{Universal compressive tomography in the time-frequency domain}  
\author[1]{Jano Gil-Lopez}
\author[2,*]{Yong Siah Teo}
\author[1]{Syamsundar De}
\author[1]{Benjamin Brecht}
\author[2]{Hyunseok Jeong}
\author[1]{Christine Silberhorn}
\author[3,4]{Luis L. S\'{a}nchez-Soto}

\affil[1]{Integrated Quantum Optics Group, Institute for Photonic Quantum Systems (PhoQS), Paderborn University, Warburger Stra\ss e 100, 33098 Paderborn, Germany}
\affil[2]{Department of Physics and Astronomy, Seoul National University, 08826 Seoul, Korea} 
\affil[3]{Departamento de \'Optica, Facultad de F\'{\i}sica, Universidad Complutense, 28040 Madrid, Spain} 
\affil[4]{Max-Planck-Institut f\"ur die Physik des Lichts, Staudtstra\ss e 2, 91058 Erlangen, Germany}

\affil[*]{Corresponding author: \href{mailto:ys\_teo@snu.ac.kr}{ys\_teo@snu.ac.kr}}

\begin{abstract}
We implement a compressive quantum state tomography capable of reconstructing any arbitrary low-rank spectral-temporal optical signal with extremely few measurement settings and without any \emph{ad hoc} assumptions about the initially unknown signal.  This is carried out with a quantum pulse gate, a device that flexibly implements projections onto arbitrary user-specified optical modes. We present conclusive experimental results for both temporal pulsed modes and frequency bins, which showcase the versatility of our randomized compressive method and thereby introduce a universal optical reconstruction framework to these platforms.
\end{abstract}

\setboolean{displaycopyright}{true}

\begin{document}

\maketitle

\section{Introduction}

Encoding quantum information in time and frequency domains~\cite{Brecht:2015aa,Harder:2017aa,Sergei:2019qip} has gained significant attention and has been proven to be a suitable alternative for scalable quantum information processing~\cite{Gisin2007,OBrien:2009fk,Pirandola2018}. These encodings allow one to access high-dimensional Hilbert spaces, which may provide enhancements to quantum information extraction, cryptography, and communication tasks~\cite{Zhou:2003a,Lanyon2009,Babazadeh:2017aa,Silberhorn:2017aa,Cozzolino:2019aa,Yin:2020aa,Raymer:2020aa}. In addition, such encodings distinghuish themselves by being directly compatable with single-mode fiber networks, because they occupy only one single spatial mode. However, reliable time measurements with high sufficient resolution are still challenging, in particular at telecommunication wavelengths.

Achieving the quantum performance in applications requires an efficient and trustworthy characterization of the experimental procedures, which is  the scope of tomography. The proper and unambiguous estimation of quantum states with minimal resources is thus of paramount importance. The infinite-dimensional Hilbert space describing these encodings demands a computationally effective and experimentally feasible procedure. 

Compressive schemes have been concocted to efficiently reduce the measurement settings required to reconstruct a signal~\cite{Candes:2006rm}. However, they require a precise knowledge of the maximal rank of the unknown state, which is not always feasible in realistic scenarios. To bypass this drawback, in Refs.~\cite{Ahn:2019aa,Ahn:2019ns,Kim:2020aa,Teo:2020cs,Gianani:2020aa}, new compressive schemes have been designed to characterize various low-rank states, gates, and measurements in different degrees of freedom.  Crucially, they require no assumptions about the unknown quantum objects in question. 

In this work, we develop and experimentally implement a compressive tomography in the time-frequency (TF) domain that allows us to uniquely determine unknown signal states that are near-coherent using very few measurement configurations. This is especially relevant in the single-photon regime, where a small number of copies of a state and practical limitations on measurement times require the efficient use of resources. 

A critical component for our goal is the quantum pulse  gate~(QPG)~\cite{Eckstein:2011aa,Brecht:2011aa,Brecht_2011}, which can perform projections of a random input on tailored time-frequency modes. It is fed by spectrally shaped gating pulses to select time-frequency modes from the input. By shaping the gating pulse into all modes from a selected basis, one can fully scan a random input in the basis. We stress that the QPG operates on superpositions of time and/or spectral components. The QPG is already a well-stablished device for projective measurements on the temporal domain and its complete tomography has been performed~\cite{Ansari:2017c}.

We shall first introduce the basic elements concerning the kinematic description of the electromagnetic field in the TF domain and the compressive state tomography applied to arbitrary near-coherent (low-rank) signal states encoded in this domain. After detailing the experimental techniques, we present a novel set of compressive tomography results for signals encoded in two broad classes of specral-temporal formats, namely the TF pulsed modes~\cite{Iaconis_2000,Morin:2013aa,Ansari:2017aa} that encode both temporal and spectral information, and frequency bins~\cite{Lukens:2017aa,Riel_nder_2017,Lu:2018aa,Lukens:2019aa,Lu2019,Chuprina:2019aa,Lu:2020aa} that reflect solely the spectral content. These results show that compressive characterization of arbitrary optical states is feasible and achievable with only   a meager number of measurement settings.

\section{{Photon time-frequency modes}}
\label{sec:TFmod}

\subsection{Kinematics}
Although all the concepts we will use here apply to any state of light, we restrict our attention  to single-photon states, as they  constitute our main optical sources for all experimental demonstrations. For a fixed polarization and transverse field distribution, a single-photon quantum state can be expressed as the coherent superposition
\begin{equation}  
|\psi \rangle = \int \D \omega\,\widetilde{u}(\omega) \, 
a_{\omega}^\dag \ket{0} \, , 
\end{equation}
where  $a^{\dagger}_{\omega}$ is the standard creation operator and $\widetilde{u} (\omega)$ is the complex spectral amplitude of the wave packet. The state can alternatively be expressed as
\begin{equation}
|\psi \rangle = \int \D t \, u(t) \, a_{t}^\dag \ket{0}\, ,
\end{equation}
where the mode functions in the respective domains are Fourier transforms of each other. Such a state belongs to the infinite-dimensional Hilbert space spanned by the continuous bases $\{\ket{t} \equiv a^\dag_{t}\ket{0}\}$ and $\{ \ket{\omega} \equiv a^\dag_{\omega}\ket{0}\}$. Both bases allow for a resolution of the identity
\begin{equation}  
\int\D t \, \ket{t} \bra{t} = \openone = 
\int \D \omega \,\ket{\omega}\bra{\omega} \, , 
\end{equation}
and their overlap is $\inner{t}{\omega}=\E{\I t\omega}/\sqrt{2\pi}$, confirming that they are conjugate variables.

One can easily extend this formalism to describe mixed states in the form 
\begin{align}
\varrho & = \int\D t'\,\int\D t''\,\ket{t'}\,u(t',t'')\,\bra{t''}\nonumber\\
& = \int\D \omega'\,\int\D \omega''\,\ket{\omega'}\,\widetilde{u}(\omega',\omega'')\,\bra{\omega''}\,\geq 0\, .
\label{eq:rho}
\end{align}
Hermiticity imposes $u^{\ast}(t',t'') = u(t'',t')$ and $\widetilde{u}^{\ast}(\omega',\omega'') = \widetilde{u}(\omega'',\omega')$, whereas positivity gives
\begin{equation}
\int\D t \, u(t,t) = 1 = \int \D \omega \, \widetilde{u} (\omega, \omega) \, . 
\end{equation}

Since $\varrho$ encodes all information accessible in both domains, it is in principle possible to explore the full spectral-temporal content of a given signal by placing both time and frequency on equal footing. The most suitable way to do this is by using the so-called chronocyclic Wigner function~\cite{Paye:1992aa,Beck:1993aa,Praxmeyer:2005aa,Walmsley:2009aa,Brecht:2013aa}, which is an adapted version of the original idea of Wigner~\cite{Wigner:1932aa}. It is defined as
\begin{align}
W(t,\omega)=&\,2\int\D t'\,\E{2\I \omega t'}\,\opinner{t-t'}{\varrho}{t+t'}\,,\nonumber\\
=&\,2\int\D \omega'\,\E{-2\I \omega' t}\,\opinner{\omega-\omega'}{\varrho}{\omega+\omega'}\,,
\end{align}
with normalization  $\int\D t'\D\omega'\,W(t',\omega')/(2\pi)=1$. The term ``chronocyclic'' signifies the juxtaposed appearances of both time and frequency variables.

A convenient TF encoding can be accomplished through basis projections~\cite{Ansari:2020aa}. More specifically, by employing a finite set of states $\sum^{d-1}_{n=0}\ket{\phi_n} \bra{\phi_n}= \openone_{d}$ that span a $d$-dimensional subspace, we may define the $d$-dimensional state 
\begin{equation}
\varrho_{d} = \sum^{d-1}_{n, n'=0} \ket{\phi_n} \, 
\opinner{\phi_n}{\varrho}{\phi_{n'}} \, \bra{\phi_{n'}}\, .
\end{equation} 
A handy basis to express this finite-dimensional state is the set of Hermite-Gaussian~(HG) modes~\cite{Patera:2018aa,Ansari:2018he,Ansari:2018to,Cox:2019aa} ($\ket{\phi_n}\equiv\ket{\mathrm{HG}_{n}}$):
\begin{equation}
\inner{t}{\mathrm{HG}_n}= \frac{1}{\sqrt{\pi^{1/2} 2^{n} n!}}
\E{-t^2/2}\,\HERM{n}{t} \, , 
\end{equation} 
where $\HERM{n} (t)$ is the Hermite polynomial. In these modes, the Wigner function is given by~\cite{Teo:2012in,Teo:2015qs}
\begin{align}
W(t,\omega) & = 2 \,\E{-2\,|\alpha|^2} \sum^{d-1}_{n, n'=0} 
(-1)^{n_<}\, \varrho_{nn'} \,\dfrac{2^{n_>}}{2^{n_<}}
\sqrt{\dfrac{n_<!}{n_>!}}\nonumber\\
& \times \,|\alpha|^{n_>-n_<}\,\E{-\I(m-n)\theta}\,\ALAG{n_<}{n_>-n_<}{4|\alpha|^2}\,,
\label{eq:wig_fock_rep}
\end{align}
where $\varrho_{nn'}$ are the matrix elements of $\varrho$ in that basis, $\ALAG{\mu}{n}{y}$ is the associated Laguerre polynomial, $\alpha=(t+\I\omega)/\sqrt{2}=|\alpha|\,\E{\I\theta}$, $n_>=\max\{n,n' \}$, and $n_<=\min\{n,n' \}$. For the HG modes, the Wigner function reduces to
\begin{equation}
W_n(t,\omega)= 2 \,(-1)^n\,\E{-(t^2+\omega^2)^2}\,\LAG{n}{2t^2+2\omega^2} \,, 
\label{eq:wig_fock}
\end{equation} 
that is, simple Laguerre-Gaussian functions.

Pulse shapers~\cite{Ansari:2018aa} can be used to generate pulsed modes of arbitrary spectral-temporal content. As a special case, one can fashion pulses that are well-enveloped only in the frequency domain. A specific type of such pulses are frequency bins, where discrete spectra of narrowband frequencies are selected to define their single-mode states. From the spectral decomposition (\ref{eq:rho}), the mode amplitude $\widetilde{u}(\omega',\omega'')$ of these frequency bins takes the ideal form
\begin{equation}
\widetilde{u}(\omega',\omega'')= \sum^{d-1}_{n, n'=0}
\widetilde{u}_{n\,n'}\,\delta(\omega'-\omega_n)\,
\delta(\omega''-\omega_{n'}) \, ,
\label{eq:freq-bin_rho}
\end{equation}
for a set of $d$ frequency bands $\{\omega_n\}^{d-1}_{n=0}$. The resulting $d$-dimensional frequency-bin state derived from the second equality in \eqref{eq:rho},
\begin{equation}
\varrho= \sum^{d-1}_{n, n'=0} \ket{\omega_n}\widetilde{u}_{n\,n'}
\bra{\omega_{n'}} \, ,
\end{equation}
thus has the Wigner function
\begin{align}
W(t,\omega) & = \sum^{d-1}_{n, n'=0} 
\widetilde{u}_{n\,n'}\,\E{\I(\omega_n-\omega_{n'})t}\,
\delta(\omega-(\omega_n+\omega_{n'})/2) \, .
\end{align}
Although $W(t,\omega)$ is now plane-wave oscillatory in $t$ and singular in $\omega$, as is expected from an overidealized model that has perfectly well-defined frequency bands of zero width in the spectral domain, terms in $t$ and $\omega$ may be disregarded as they are auxiliary. Operationally, it is the complex amplitudes $\widetilde{u}_{n\,n'}$ that truly contain all purely-spectral information about the frequency-binned system, and can hence be easily represented by their positive matrix $\widetilde{\dyadic{u}}=\sum^{d-1}_{n, n'=0} \rvec{e}_n\widetilde{u}_{n\,n'}\TP{\rvec{e}_{n'}}$ in some computational basis $\{\rvec{e}_n\}$.

\subsection{Generalized measurements with the quantum pulse gate}
\label{subsec:qpg}

\begin{figure}[t]
    \centering
    \includegraphics[width=\columnwidth/2]{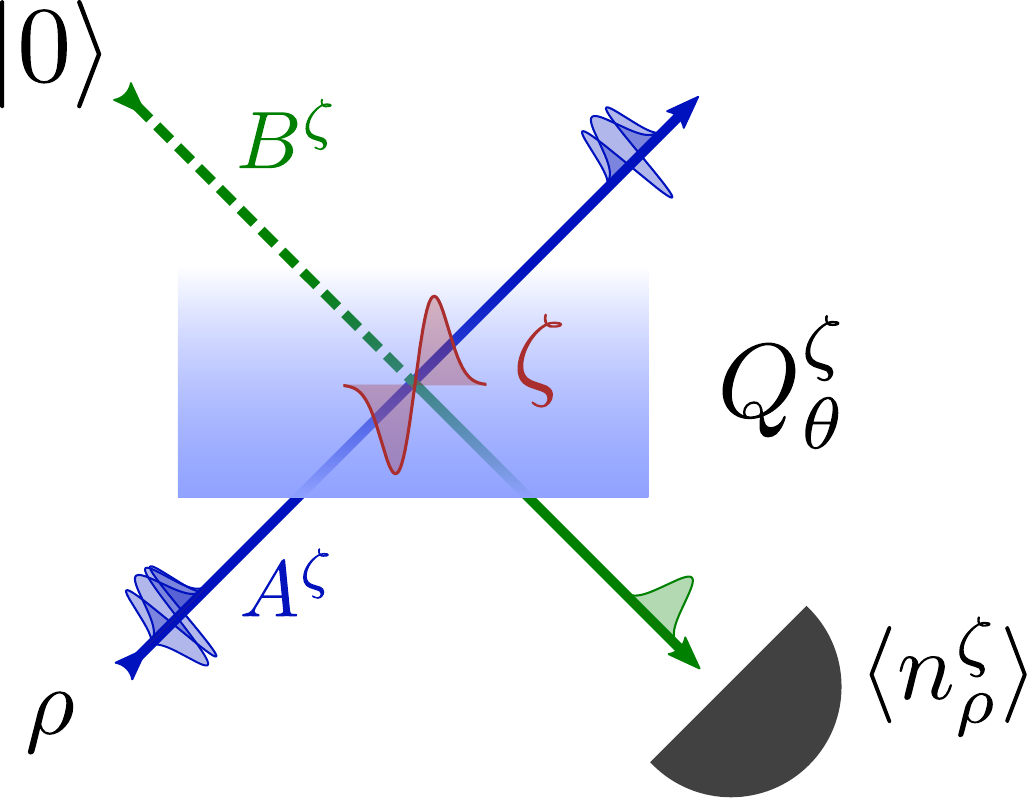}
    \caption{\label{fig:QPG}~The QPG implements a beam splitter operation between two sets of modes $\{A\}$ and $\{B\}$, where one user chosen input mode $A^\zeta$ is converted to an output mode $B$, while all other modes are transmitted. Photon detection in the output mode then implements a projection of an input state $\varrho$ onto mode $A^\zeta$.} 
\end{figure}

An ideal QPG mode matched to the source acts on an arbitrary single-photon input state $\varrho_{\mathrm{in}}$ according to~\cite{Brecht:2015aa}
\begin{equation}
\varrho_{\mathrm{out}} = Q^{\zeta}_{\theta} \, \varrho_{\mathrm{in}} \, Q^{\zeta \, \dagger}_{\theta} .
\end{equation}
where
\begin{align}
    Q^\zeta_\theta & = \openone - \ket{A^\zeta}\bra{A^\zeta} -
    \ket{B}\bra{B} \nonumber \\
    & + \cos{\theta} \left( \ket{A^\zeta}\bra{A^\zeta} +  
    \ket{B}\bra{B} \right) \nonumber \\
    & + \sin\theta \left(  \ket{B}\bra{A^\zeta} - \ket{A^\zeta}\bra{B} \right),
\end{align}
where 
\begin{equation}
\ket{A^\zeta} = A^{\zeta \, \dagger} \ket{0} \equiv 
\int \D\omega \zeta(\omega)a^\dagger_\omega \ket{0}
\end{equation} 
is the source mode. This consists of a family of unitary transformations on the single-photon state space composed of two nonoverlapping frequency bands: one spanned by the input mode state $\ket{A^\zeta}$, and  a single TF mode $\ket{B}$ occupying the other. Note that this operation describes a special quantum mechanical beam splitter, as sketched in Fig.~\ref{fig:QPG}.

Given an input state with photon mean number $N_\varrho$, the mean photon number in the output mode $\ket{B}$ of the QPG is
\begin{equation}
    \langle N^\zeta_\varrho\rangle = N_\varrho \, \eta\, |\gamma^\zeta_\varrho|^2,
\end{equation}
with $\eta= \sin^2\theta$ the conversion efficiency of the QPG and $\gamma^\zeta_\varrho = \bra{A^\zeta}\varrho\ket{A^\zeta}$ the overlap between the input mode $\ket{A^\zeta}$ of the QPG and the input state $\varrho$.

Only the part of $\varrho$ overlapping with the QPG input mode is converted to mode $\ket{B}$. Subsequent photon counting in this mode then effectively implements a projection of $\varrho$ onto $\ket{A^\zeta}$, where $\ket{A^\zeta}\bra{A^\zeta}$ is a projector of a measurement basis. 

\section{Randomized compressive tomography}

A rank-$r$ state of dimension $d$, described by a positive unit-trace operator $\varrho_d$, can be uniquely determined by $(2d-r)r-1$ independent parameters. On the other hand, a set of $\lceil[(2d-r)r-1]/(d-1)\rceil$ measurements generally cannot fully characterize such a state. For instance, it has been shown that two bases can never fully characterize arbitrary states of $d=2$---the finite version of the Pauli problem~\cite{Pauli:1933pr,Peres:1993aa}. Nevertheless, one can still search for a set of $M\ll O(d^2)$ measurement outcomes that can unambiguously reconstruct the set of states of known rank $r\ll d$. We call such a measurement set informationally complete~(IC). However, when $r$ and all other information about $\varrho$ are unknown to the observer, surmising that a fixed set of measurements with an $M\ll O(d^2)$ would completely determine $\varrho_d$ before a tomography experiment generally leads to unreliable reconstruction results.

In this section, we introduce a randomized compressive tomography scheme~(RCT) for characterizing an unknown low-rank time-frequency state $\varrho_d$ of a finite dimension $d$ and rank $r\ll d$. We show that without resorting to any \emph{ad hoc} assumption about $\varrho_d$, we can still determine whether a given number $M\ll O(d^2)$ of randomly-chosen orthonormal bases can uniquely reconstruct $\varrho_d$. Each basis is denoted as $\mathcal{B}_{k}=\{\ket{b_{k0}},\ket{b_{k1}},\ldots,\ket{b_{k\,d-1}}\}$, with $\sum^{d-1}_{l=0}\ket{b_{kl}}\bra{b_{kl}}= \openone$. This scheme is therefore universal, in the sense that with it, states of arbitrary time-frequency modes can be compressively characterized using general bases measurements that can be very reliably generated using the QPG, as discussed in Sec.~\ref{sec:TFmod}.\ref{subsec:qpg}.

\begin{figure}[t]
    \centering
    \includegraphics[width=1\columnwidth]{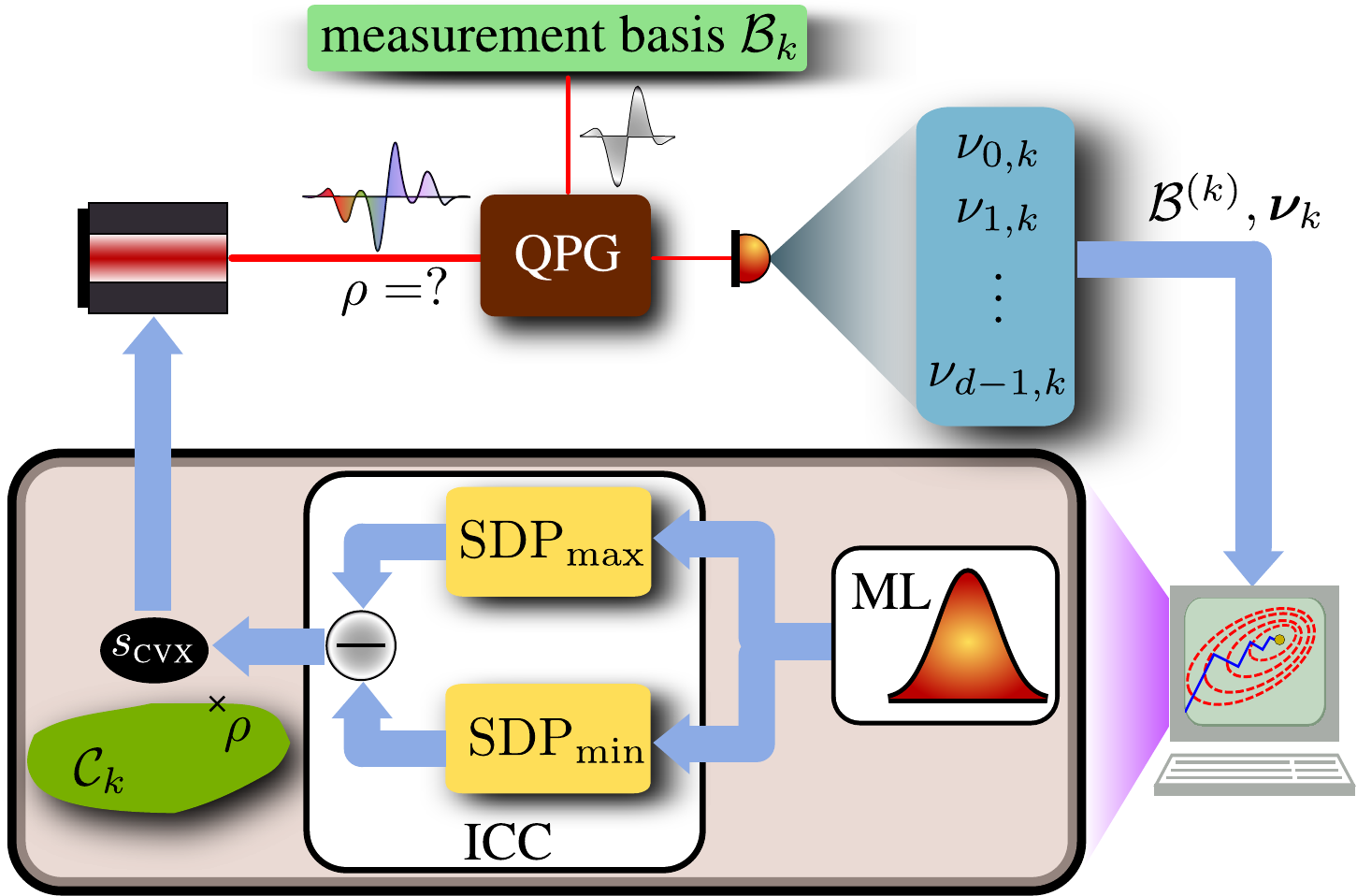}
    \caption{\label{fig:RCT}Schematic figure of the iterative RCT scheme. The signal carrying the unknown state $\varrho$ interacts with the QPG so that a randomly-chosen basis is measured in the $k$th step. This gives a set of relative frequencies that is combined with previous measurements. All $k$ measured bases $\mathcal{B}^{(k)}$ and their corresponding relative frequencies $\rvec{\nu}_k$ are then processed numerically by first carrying out the ML routine to obtain physical probabilities, and next subjecting the results to the ICC algorithm that computes the completeness indicator $s_\textsc{cvx}$ using two semidefinite programs (SDPs)~(discussed in Appendix~\ref{app:ICC}). The whole cycle repeats until $s_\textsc{cvx}$ drops below a certain small threshold at $k=K_{\textsc{ic}}$, implying that $(\mathcal{B}^{(K_{\textsc{ic}})},\rvec{\nu}_{K_{\textsc{ic}}})$ is IC.}
\end{figure}

The RCT scheme is a bottom-up iterative procedure, in which independently chosen measurement bases $\mathcal{B}^{(k)}=\{\mathcal{B}_1,\mathcal{B}_2,\ldots,\mathcal{B}_k\}$ are measured and accumulated until the time-frequency state estimator $\widehat{\varrho}_d\geq0$ is unique. When this happens, it implies that apart from $\widehat{\varrho}_d$, there is no other state that is consistent with the measurements. More specifically, in the $k$th iterative step, after a basis $\mathcal{B}_k$ is measured, the accumulated bases set $\mathcal{B}^{(k)}$ and corresponding relative frequency data $\rvec{\nu}_k=(\nu_{10},\nu_{11},\ldots,\nu_{1\,d-1},\ldots,\nu_{k0},\nu_{k1},\ldots,\nu_{k\,d-1})^{\top}$ for these bases ($\sum^{d-1}_{l=0}\nu_{kl}=1$) are analyzed to see if these measurements are IC. 

This procedure involves two following stages as illustrated in Fig.~\ref{fig:RCT}. In the first stage, the column of raw relative frequencies $\rvec{\nu}_k$ are mapped to the corresponding column $\widehat{\rvec{p}}_k$ whose elements $\widehat{p}_{kl}=\opinner{b_{kl}}{\varrho_d}{b_{kl}}$ are physical probabilities obtained from some positive unit-trace operator $\varrho_d$. This mapping is necessary to ensure that the analysis is physical. For this, we may invoke the maximum-likelihood~(ML) method~\cite{Aldrich:1997ml,Banaszek:1999ml,Rehacek:2007ml,Teo:2011me,Shang:2017sf} that would give us the column $\widehat{\rvec{p}}_k$ that maximizes the likelihood function describing the QPG measurement scenario over the physical $d$-dimensional state space~\cite{github:2021}.

The second stage of RCT at the $k$th step is to find out if there is more than one state that gives such physical ML probabilities $p_{jk}$---the informational completeness certification~(ICC). If this were true, then in principle, there will be convex set ($\mathcal{C}_k$) of states with a nonzero volume. The task is to deterministically figure out the value $k=K_\textsc{ic}$ at which the volume of $\mathcal{C}_{K_\textsc{ic}}$ is zero. To do this, we introduce an indicator $s_{\textsc{cvx}}$ that monotonically decreases with the convex-set volume. When $s_{\textsc{cvx},K_\textsc{ic}}=0$, it can be argued easily that $\mathcal{C}_{K_\textsc{ic}}$ is a single point, telling us that ($\mathcal{B}^{(K_\textsc{ic})},\rvec{\nu}_{K_\textsc{ic}}$) is IC~\cite{Ahn:2019aa}. The computation of $s_\textsc{cvx}$ can be done with the help of semidefinite programming and is explained in Appendix~\ref{app:ICC}.

\begin{figure*}[t]
    \centering
    \includegraphics[width=\textwidth]{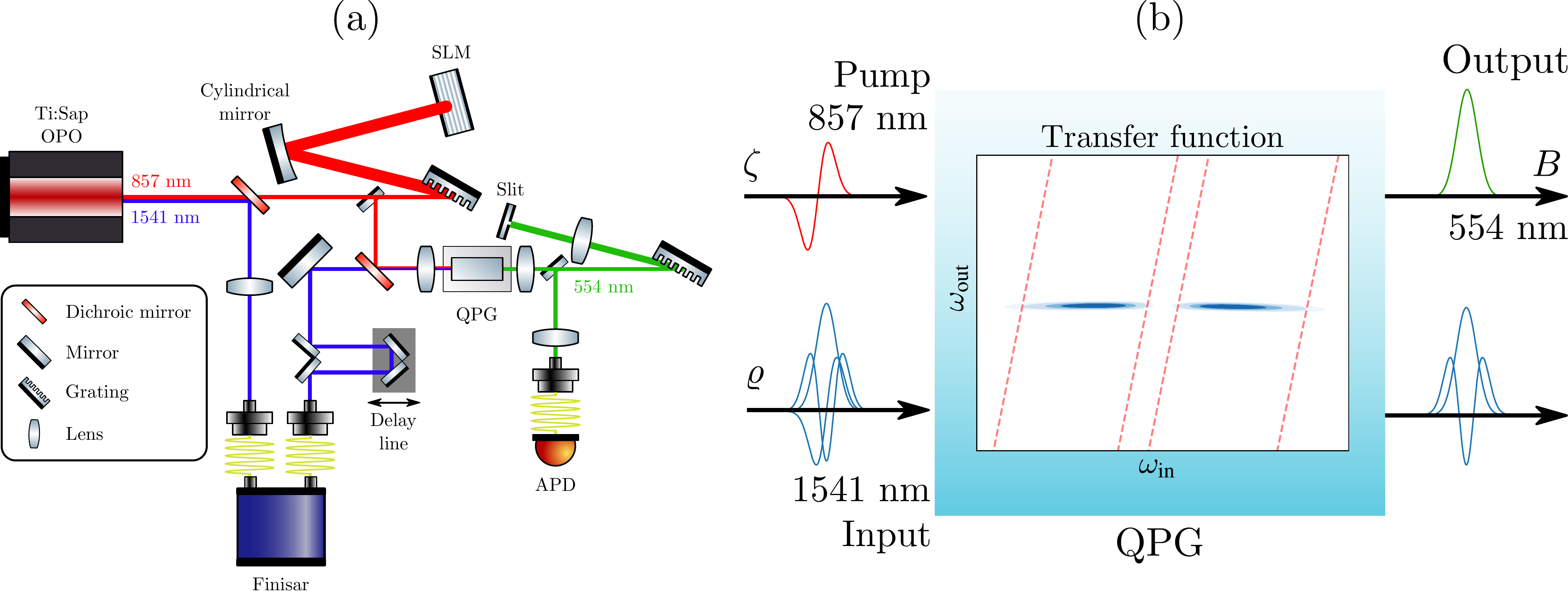}
    \caption{(a) Experimental setup. A Ti:sap-OPO laser system emits ultrafast pulses at the required wavelengths. The input signal at telecommunication wavelengths is spectrally shaped using a comercial shaper and the pump field is shaped using a home-built $4f$ line shaper setup with an SLM device. A delay line on the input field arm is used to match it temporally with the pump field for the QPG process. At the output of the QPG, unconverted input and throughput pump are separeted from the output field with dichroic mirrors that were not depicted to simplify the sketch. The output field is then filtered with a home-made spectral filter and sent to the APD. (b) Experimental realisation of the QPG. Input and pump fields are coupled into the QPG waveguide device where the process' transfer function selects the mode $A^\zeta$ from the input state $\varrho$ and upconverts it to the output mode $B$. The transfer function $\varphi(\omega_{\mathrm{in}}, \omega_{\mathrm{out}})$ inside the QPG describes the relation betwen the input and output frequencies depending on the phasematching function $\phi$ and the pump mode $\zeta$. The pump amplitude at $1/\text{e}^2$ is plotted in red dashed lines as a reference. }
    \label{fig:setup}
\end{figure*}

\section{Experimental techniques}

In our realization, we implement the QPG with a group-velocity matched (GVM) sum-frequency generation in a periodically poled, 35~mm long and 7 \textmu m wide titanium in-diffused lithium niobate waveguide. The 4.4 \textmu m poling period grants quasi-phasematching at the desired GVM wavelengths; 1541~nm for the input state $\varrho$ and 857~nm for the pump mode $\zeta$. The GVM process is engineered so that the output mode $\ket{B}$ at 554~nm is solely defined by the phasematching function $\phi (\omega_{\mathrm{in}}, \omega_{\mathrm{out}})$ of the process. This realizes the mode-selective process described in Sect.~\ref{sec:TFmod}.\ref{subsec:qpg}.

The transfer function is then the product of the phasematching function and the pump mode envelope $\varphi( \omega_{\mathrm{in}}, \omega_{\mathrm{out}})=\zeta(\omega_{\mathrm{pump}}) \phi(\omega_{\mathrm{in}}, \omega_{\mathrm{out}})$. It describes the relation between input and output frequencies for the SFG process. An example of a transfer function with a first-order Hermite-Gaussian envelope of the pump mode is depicted in the inset in Fig.~\ref{fig:setup} (b). Only the parts of the input state that overlap with the transfer function will be converted to the output and detected with a single-photon detector. Hence, for any input state, the QPG performs a projective measurement on the mode defined by the pump. 

The input field is shaped into the appropriate input states $\varrho_{\mathrm{in}}$ using a fiber-coupled commercial spectral shaper (Finisar Waveshaper). The output of the shaper is then coupled to the QPG in free space with a lens. A delay line is used to match the arrival time of the signal pulses to those of the pump at the QPG. 
The pump field is sent to a home-built spectral shaper setup consisting of a holographic grating (2000~lines/mm), a cylindrical mirror and a Spatial Light Modulator (SLM) (Hamamatsu LCOS-SLM X12513-07) in a folded $4f$ line configuration. This allows to shape the pump envelope $\zeta$ into any base component for the RCT measurements. The shaped pump pulses are then coupled to the QPG through the same optical path as the input field.

At the output of the QPG, the unconverted input and transmitted pump fields are separated from the up-converted output with dichroic mirrors. The up-converted green output field is then sent to a tunable spectral filter consisting of a grating, a lens and a slit with a variable width with a mirror on its back, in a folded 4f line configuration. The filter is set to 30~pm FWHM to filter out the side lobes of the QPG $\sinc$ phase-matching function~\cite{Ansari:2018aa} and to further increase the selectivity of the mode-selective process. The filtered output is then sent to an avalanche photon detector (APD) (ID Quantique) connected to a time-tagger (Swabian instruments) to collect the measurement data.

The input states to be reconstructed are chosen to be either temporal modes whose envelopes are HG functions or frequency bins of dimension~$d$. The temporal modes have an spectral FWHM of 1~THz, the frequency bins are 0.07~THz wide. The pump modes are then shaped into the $d$-dimensional randomly rotated basis modes $\mathcal{B}_k$ of each input state. Note that RCT exhibits compressive effects with any pump modes, since the ICC procedure always decisively verify whether any given measurement dataset is IC regardless of the unknown state. The rotation uses randomly generated unitary matrices.

\begin{figure*}[t]
    \centering
    \includegraphics[width=1.96\columnwidth]{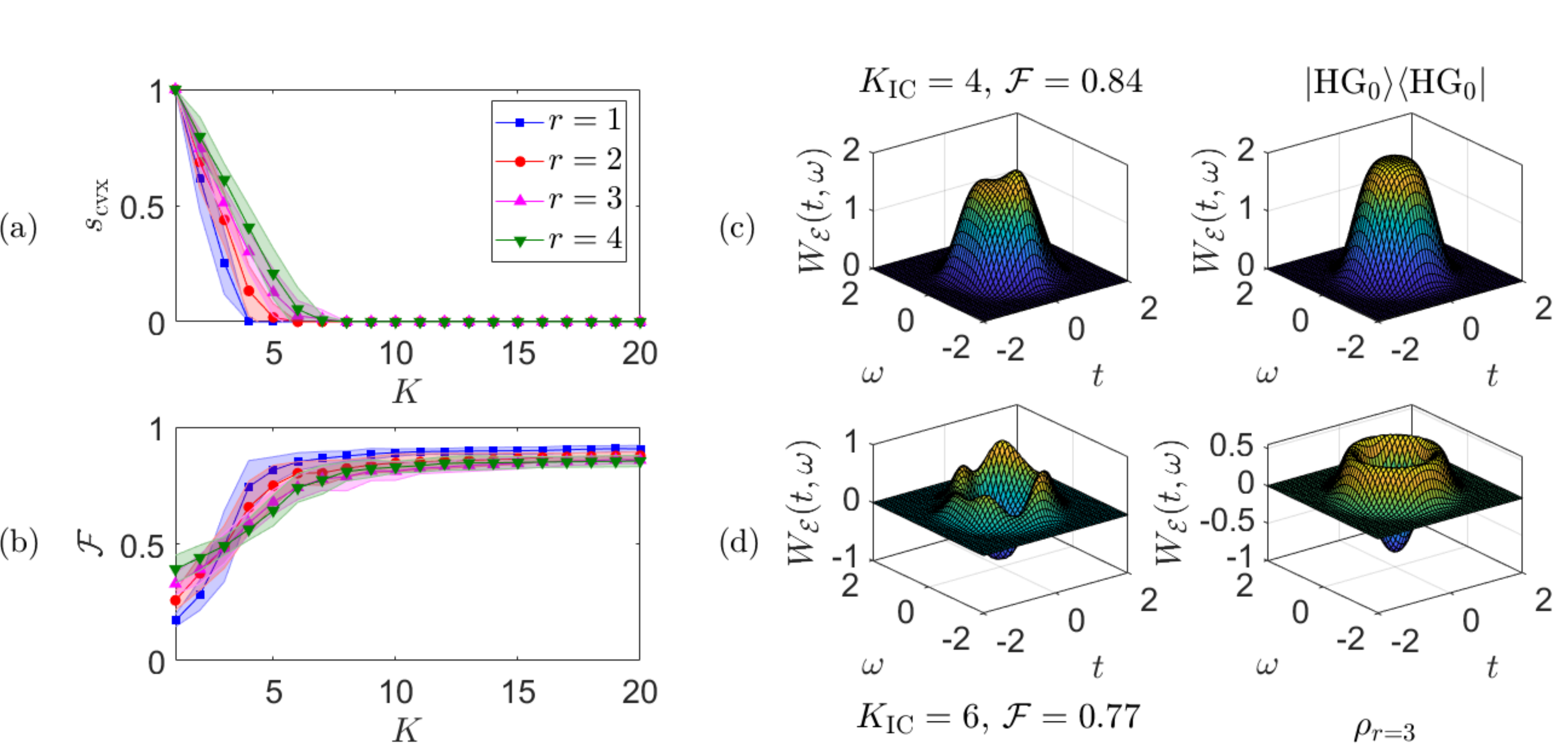}
    \caption{\label{fig:resa}Reconstruction results for time-frequency states of $d=10$. (a,b)~The average drop in $s_\textsc{cvx}$ for each state rank $r$ coincides with the average rise in fidelity $\mathcal{F}$. The 1-$\sigma$ error regions are computed with 10 experimental runs per value of $r$. The value of $K_\textsc{ic}$ for which $s_{\textsc{cvx},K_\textsc{ic}}=0$ steadily increases with $r$ as it should. Whenever $s_\textsc{cvx}\neq0$, $\mathcal{F}$ is computed for $\varrho_\text{min}$ (see Appendix~\ref{app:ICC}). (c,d)~Sample Wigner functions for the rank-one HG$_0$ mode and a random rank-3 mixture of the HG modes ($\varrho_{r=3}=\ket{\mathrm{HG}_0}0.17\bra{\mathrm{HG}_0}+\ket{\mathrm{HG}_1}0.70\bra{\mathrm{HG}_1}+\ket{\mathrm{HG}_2}0.13\bra{\mathrm{HG}_2}$) are shown.}
\end{figure*}

For every input state under investigation, the input field is projected onto each basis mode of the $M$ different randomly rotated bases in individual measurements. The resulting counts $\nu_{jk}$ at the output $B$ are measured and stored. Every 10 measurements, the time delay between signal input and pump is realigned automatically by maximizing the mode-selectivity to account for drifts in the setup. To realize measurements on mixed input states, we measure different inputs with the same set of randomly rotated bases. We then mix the measured data with appropriate weights in post-processing. The randomly rotated bases implement a non-uniform sampling of the input state that accomplishes compressive sensing \cite{Gibson2020, Tomm2021}. In a sense, the QPG can be considered a single-pixel camera for temporal modes. States with temporal features much faster than the resolution of the single-photon detector are reconstructed, by modulating the input signal with random temporal masks. In addition, the use of compressed sensing facilitates signal reconstruction with fewer measurements than a direct sampling approach~\cite{Allgaier:2017c}. This is especially beneficial in situations in which a signal must be reconstructed from a limited number of photons.

\section{Results}
\label{sec:res}

\subsection{Time-frequency modes}

The first class of low-rank TF states we shall use to demonstrate the RCT scheme constitutes the HG~modes and their statistical mixtures. These are highly relevant as they closely approximate the eigenbasis of parametric down-conversion processes~\cite{Law:2000jo}and have been shown to be optimal for certain metrology tasks. In terms of their chronocyclic Wigner function representation, states corresponding to the first three HG modes, for instance, are expressed as
\begin{align}
    W_{n=0}(t,\omega)=&\,2\,\E{-y^2}\,,\nonumber\\
    W_{n=1}(t,\omega)=&\,2\,\E{-y^2}\,(2y-1)\,,\nonumber\\
    W_{n=2}(t,\omega)=&\,2\,\E{-y^2}\,(2y^2-4y+1)\,,
\end{align}
where $y=t^2+\omega^2$. These states are all rotationally symmetric in the spectral-temporal content.

A total of ten HG basis modes, $\sum^{9}_{n=0}\ket{\mathrm{HG}_n}\bra{\mathrm{HG}_n}=\openone_{10}$, are used to project $\varrho$ onto a finite-dimensional subspace of dimension $d=10$. Random von~Neumann basis measurements of the same dimension are generated with the QPG to collect measurement data for the rank-one modes HG$_0$, HG$_1$, HG$_2$ and HG$_3$, and their statistical mixtures. These basis measurements are parametrized by random unitary rotations distributed according to the Haar measure~\cite{Mezzadri:2007qr}. Figure~\ref{fig:resa} shows the results for the different low-rank time-frequency states. For the rank-one graphs in panels~(a) and (b), the results are averaged over experimental runs of all the four HG modes.

As the number of independent parameters characterizing a rank-$r$ state increases roughly linearly in $r$ for $r\ll d$, the number of measurement bases needed to uniquely reconstruct rank-$r$ states should also increase roughly linearly in $r$ in this regime~\cite{Ahn:2019ns}. In other words, on average, the number of bases needed to fully characterize rank-$r$ time-frequency states is $O(r)$. Such a characteristic can be observed in Figs.~\ref{fig:resa}(a) and (b), despite the presence of statistical noise and experimental imperfections in the measurement data. The results therefore demonstrate the efficiency and robustness of RCT when applied to real experimental scenarios. Chronocylic Wigner functions of sample reconstructions are also shown in Figs.~\ref{fig:resa}(c) and (d) for visualization.

\begin{figure*}[t]
    \centering
    \includegraphics[width=2\columnwidth]{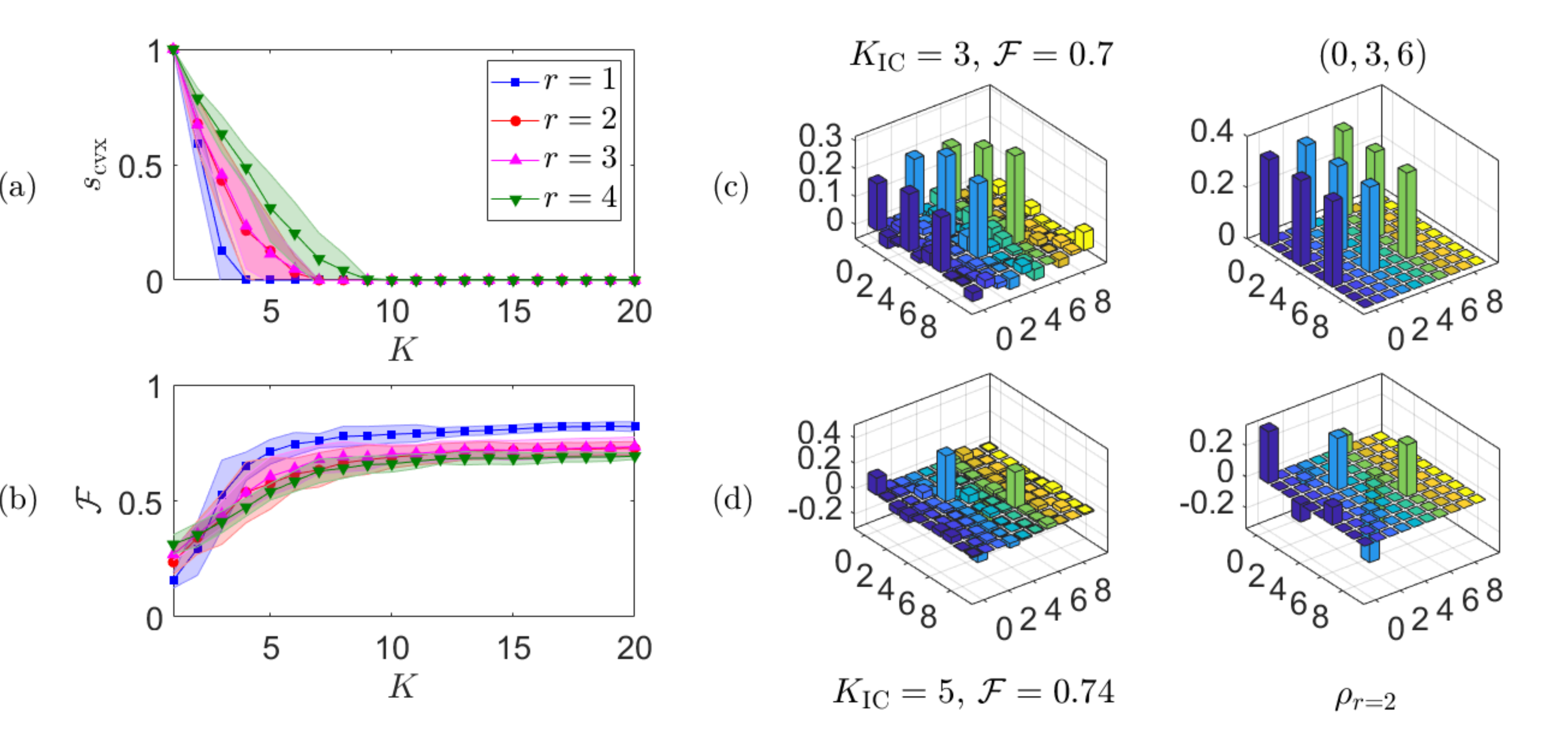}
    \caption{\label{fig:resb}Reconstruction results for frequency-bin states of $d=10$, with technical specifications following those of Fig.~\ref{fig:resa}. (a,b) The behaviors of $s_\textsc{cvx}$ and $\mathcal{F}$ for each rank of $\varrho$ or $\widetilde{\dyadic{u}}$ follows similarly to those in Fig.~\ref{fig:resa}. (c,d)~Sample matrix plots of $\RE\{\widetilde{\dyadic{u}}\}$ for a rank-one frequency-bin superposition  ($\varrho=\ket{\psi}\bra{\psi}$ with $\ket{\psi}=\ket{\omega_0}+\ket{\omega_3}+\ket{\omega_6}/\sqrt{3}$) and a random rank-2 mixture of frequency bins (described by a $\widetilde{\dyadic{u}}$ with eigenvalues 0.73 and 0.27) are shown.}
\end{figure*}

\subsection{Frequency bins}

The second class of low-rank states that we consider here are frequency-bin states. These are states defined by a discrete set of narrowband frequency bins. The corresponding complex matrix $\widetilde{\dyadic{u}}$ that dresses these continuous frequency bases inasmuch as Eq.~\eqref{eq:freq-bin_rho} is a positive matrix that full information about a general state defined by these frequency bins in the spectral domain. 

In particular, we focus on 10-dimensional states supported by 10 pre-chosen frequency bins $\{\ket{\omega_n}\bra{\omega_n}\}^9_{n=0}$. We generate four rank-one states $\varrho=\ket{\psi}\bra{\psi}$  that are superpositions of such bins, namely
\begin{align}
        \ket{\psi} =\begin{cases}
         (\ket{\omega_0}+\ket{\omega_3}+\ket{\omega_6})/\sqrt{3} \\
         (\ket{\omega_0}-\ket{\omega_3}+\ket{\omega_6})/\sqrt{3}\\
         (\ket{\omega_3}-\ket{\omega_6}+\ket{\omega_0})/\sqrt{3}\\
         (\ket{\omega_6}-\ket{\omega_0}+\ket{\omega_3})/\sqrt{3}
    \end{cases}.
\end{align}
For mixed states of higher ranks, we consider random statistical mixtures of the above four superpositions with random mixture probabilities. The measurements used to probe all these states are again random bases generated by unitary operators sampled uniformly from the Haar measure. 

Figure~\ref{fig:resb} gives the performances of RCT on the class of frequency-bin states. The general behavior of $s_\textsc{cvx}$ as a function of the number of bases measured, $K$, is consistent with that for the time-frequency states, confirming the basic understanding that compressive methods are system independent. The $K_\textsc{ic}$ values for $r>1$ are on average larger than the values in Fig.~\ref{fig:resa} owing to more significant experimental noise present in these states, as commensurately reflected in the fidelity graphs of Fig.~\ref{fig:resb}(b). The matrix plots of $\RE\{\widetilde{\dyadic{u}}\}$ that represents all frequency-bin states present example reconstructions one would expect in typical experiments.

\section{Conclusions}

We have demonstrated a versatile compressive quantum tomography scheme that can characterize arbitrary near-coherent quantum states in the TF domain using extremely few measurements. The method is very robust and requires no spurious assumptions about the states: this includes the degree of sparsity or coherence, that could most likely be inconsistent with the actual implementation.

From a technical perspective our method allows for the efficient characterization of the temporal behaviour of telecommunication light at the single-photon level and can thus pave the way for many new quantum technologies.

The great performance of the method largely relies on the flexibility of the QPG, which has allowed us to implement linear optics single-photon quantum operations in terms of the TF modes: the natural variables to deal with these signals in the quantum domain.  These modes are  compatible with waveguide technology, making them ideal candidates for integration into existing communication networks. In addition, they are not affected by typical medium distortions such as linear dispersion, which renders them robust basis states for real-world applications.   

Through real experimental demonstrations, we showed that our  compressive scheme can perform complete reconstruction of any TF quantum state using readily-accessible algorithms that are much more versatile than the toolkits offered by conventional coherent spectroscopy.

\begin{backmatter}
\bmsection{Funding}
National Research Foundation of Korea (NRF-2019R1A6A1A10073437, NRF-2019M3E4A1080074, NRF-2020R1A2C1008609, 2020K2A9A1A06102946),  Deutsche Forschungsgemeinschaft (SFB TRR 142), European Union's Horizon 2020 research and innovation program (ApresSF and STORMYTUNE), Ministerio de Ciencia e Innovaci{\'o}n (PGC2018-099183-B-I00).

\bmsection{Disclosures}
The authors declare no conflicts of interest

\bmsection{Data Availability Statement}
All data generated or analyzed during this study are included in this published article (and its supplementary information files). 
\end{backmatter}

\appendix

\section{Informational completeness certification}
\label{app:ICC}

Whenever a given set of bases $\mathcal{B}_{L}$ is \emph{not} IC, then by definition, there shall be (infinitely) many states corresponding to the physical ML probabilities ${\widehat{\rvec{p}}_L}$ extracted from the relative frequencies {$\rvec{\nu}_L$ of $\mathcal{B}_L$.} More precisely, there exists a convex set $\mathcal{C}_L$ of states {$\{\varrho\}$ that is consistent with the constraints $\varrho \geq0$, $\tr{\varrho}=1$ and $\opinner{b_{kl}}{\varrho}{b_{kl}}=\widehat{p}_{kl}$ for all $1\leq k\leq L$ and $0\leq l\leq d-1$.}

As $L$ increases to $L=K_\textsc{ic}$, the convex set $\mathcal{C}_L\rightarrow\mathcal{C}_{K_\textsc{ic}}$ eventually becomes a single point containing a unique estimator that is close to the unknown state provided that the number of photodetector clicks $N$ for each basis measurement is sufficiently large. The task of ICC is to determine the value of $K_\textsc{ic}$ that this happens. For this purpose, we note that since $\mathcal{C}_L$ is convex, if both the minimum and maximum values over $\mathcal{C}_L$ of a \emph{strictly convex} or \emph{strictly concave} function of $\varrho$ are equal to each other, then it must be the case that $\mathcal{C}_L$ is a single point of size 0. This argument clearly applies also to any linear function of $\varrho$.

This implies that a successful ICC involves the solution to the following equivalently semidefinite programs~\cite{Vandenberghe:1996ca}:
\begin{center}
	\begin{minipage}[c][4.5cm][c]{0.8\columnwidth}
		\noindent
		\rule{\columnwidth}{1.5pt}\\
		\textbf{ICC}\\[1ex]
		\noindent
		Maximize and minimize $f_Z(\varrho)=\tr{\varrho Z}$\\
		subject to
		\begin{itemize}
			\item $\varrho \geq0$\,,
			\item $\tr{\varrho }=1$\,,
			\item $\opinner{b_{kl}}{\varrho}{b_{kl}}=\widehat{p}_{kl}$\,.
		\end{itemize}
		Define $s_\textsc{cvx}=f^{\text{max}}_Z-f^{\text{min}}_Z$ and check if $s_\textsc{cvx}=0$.
		\rule{\columnwidth}{1.5pt}
	\end{minipage}
\end{center}
Here $Z$ is some fixed full-rank operator that is randomly chosen. This is necessary to ensure that $f_Z$ has no plateau structure that would violate the strict-convexity requirement. After obtaining the minimum $f^{\text{min}}_Z$ and maximum $f^{\text{max}}_Z$ values of $f_Z$, we may define $s_\textsc{cvx}=f^{\text{max}}_Z-f^{\text{min}}_Z$. Because of the convexity properties of the entire problem of ICC, $s_\textsc{cvx}$ turns out to be a monotonic indicator of the size of $\mathcal{C}_L$. If $s_\textsc{cvx}=0$, then $\mathcal{C}_L$ is a single point with $L\equiv K_\textsc{ic}$. In general, an analytical understanding of the behavior of $s_\textsc{cvx}$ with the number of measured bases $k$ is apparently intangible. As an alternative, we present a numerical exposition Sec.~\ref{sec:res} for the spectral-temporal optical states of interest.


\end{document}